\documentclass[showpacs]{revtex4}
\usepackage{epsfig}
\def\beq{\begin{equation}}
\def\enq{\end{equation}}
\def\beqa{\begin{eqnarray}}
\def\enqa{\end{eqnarray}}

\def\MeV{\nobreak\,\mbox{MeV}}
\def\GeV{\nobreak\,\mbox{GeV}}

\def\qq{\lag\bar{q}q\rag}

\def\sss{\lag\bar{s}s\rag}
\def\qqs{\lag\bar{s}s\rag}

\def\mixs{\lag\bar{s}g\si.Gs\rag}
\def\Gd{\lag g^2G^2\rag}
\def\G3{\lag g^3G^3\rag}

\def\rh{\rho}
\def\si{\sigma}

\def\al{\alpha}
\def\be{\beta}
\def\alma{\alpha_{max}}
\def\almi{\alpha_{min}}
\def\bemi{\beta_{min}}
\def\mme{m_{D_s^*D_{s0}^*}}

\def\lb{\label}

\newcommand{\rag}{\rangle}
\newcommand{\lag}{\langle}

\begin{document}

\title{\sc Can the $X(4350)$ narrow structure be a $1^{-+}$ exotic state?}
\author{Raphael M. Albuquerque}
\email{rma@if.usp.br}
\affiliation{Instituto de F\'{\i}sica, Universidade de S\~{a}o Paulo,
C.P. 66318, 05315-970 S\~{a}o Paulo, SP, Brazil}
\author{Jorgivan M. Dias}
\email{jdias@if.usp.br}
\affiliation{Instituto de F\'{\i}sica, Universidade de S\~{a}o Paulo,
C.P. 66318, 05315-970 S\~{a}o Paulo, SP, Brazil}
\author{Marina Nielsen}
\email{mnielsen@if.usp.br}
\affiliation{Instituto de F\'{\i}sica, Universidade de S\~{a}o Paulo,
C.P. 66318, 05315-970 S\~{a}o Paulo, SP, Brazil}

\begin{abstract}
Using the QCD  sum rules we test if the new narrow structure, the $X(4350)$
recently observed by the Belle Collaboration, can be described as a
$J^{PC}=1^{-+}$ exotic $D_s^*D_{s0}^*$ molecular state.
We consider the contributions of condensates up to dimension eight, we
work at leading order in $\alpha_s$ and we keep terms which are linear in
the strange quark mass $m_s$. The mass obtained for such state is 
$m_{D_s^*{D}_{s0}^*}=(5.05\pm 0.19)$ GeV.
We also consider a molecular $1^{-+},~D^{*}{D}_0^{*}$ 
current and we obtain $m_{D^*{D}_0^*}=(4.92\pm 0.08)$ GeV.
We conclude that it is not possible to 
describe the $X(4350)$ structure as a  $1^{-+}~D_s^*{D}_{s0}^*$ molecular 
state.
\end{abstract}

\maketitle


In the recent years, many new states were observed  by BaBar, 
Belle and CDF Collaborations. All these states were observed in decays
containing a $J/\psi$ or $\psi^\prime$ in the final states and their masses 
are in the charmonium region. Therefore, they 
certainly contain a $c\bar{c}$ pair in their constituents. Although they are
above the threshold for a decay into a pair of open charm mesons they decay
into $J/\psi$ or $\psi^\prime$ plus pions, which is unusual for $c\bar{c}$
states. Another common feature of these states is the fact that their masses 
and decay modes are not in agreement with the predictions from potential 
models. For these reasons they are considered as candidates for exotic 
states. Some of these new states have their masses
very close to the meson-meson threshold, like the $X(3872)$ \cite{belle1}
the $Z^+(4430)$ \cite{belle2} and the $Y(4140)$ \cite{cdf}. Therefore, a 
molecular interpretation for these states seems natural. 

Concerning the $Y(4140)$ structure, it was observed by  CDF Collaboration
in the decay $B^+\to Y(4140)K^+\to J/\psi\phi K^+$. The mass and width
of this structure is $M=(4143\pm2.9\pm1.2)~\MeV$, $\Gamma=(11.7^{+8.3}_{-5.0}
\pm3.7)~\MeV$ \cite{cdf}. Its interpretation as a conventional 
$c\bar{c}$ state is complicated because it lies well above the 
threshold for open charm decays and, therefore, a $c\bar{c}$ state with this 
mass would decay predominantly into an open charm pair with a large total 
width. This state  was interpreted as a $J^{PC}=0^{++}$ or $2^{++}$ 
$D_s^{*}\bar{D}_s^{*}$ molecular state in different works 
\cite{nnl,zhuy,maha,bgl,y4140,ding2,jrz,lke,oset}. In particular,
 using an effective lagrangian model, the authors of ref.~\cite{bgl} have
suggested that a $D_s^{*+}D_s^{*-}$ molecular state should be seen in 
the two-photon process. Following this suggestion the Belle Collaboration
\cite{bellex} searched for the $Y(4140)$ state in the $\gamma\gamma\to
\phi J/\psi$ process. However, instead of the $Y(4140)$, the Belle 
Collaboration found evidence for a new narrow structure in the $\phi J/\psi$
mass spectrum at 4.35 GeV. The significance of the peak is 3.2 standard 
deviations and, if interpreted as a resonance, the mass and width of
the state, called $X(4350)$ is $M=(4350.6^{+4.6}_{-5.1}\pm0.7)~\MeV$ and 
$\Gamma=(13.3^{+7.9}_{-9.1} \pm4.1)~\MeV$ \cite{bellex}.

The possible quantum numbers for a state decaying
into $J/\psi\phi$ are 
$J^{PC}=0^{++},~1^{-+}$ and $2^{++}$. At these quantum numbers, 
$1^{-+}$ is not consistent with the constituent quark model and it is 
considered exotic \cite{nnl}. In ref.~\cite{bellex} it was noted that
the mass of the $X(4350)$ is consistent with the prediction for a $cs\bar{c}
\bar{s}$ tetraquark state with $J^{PC}=2^{++}$ \cite{stancu} and a
$D_s^{*+}\bar{D}_{s0}^{*-}$ molecular state \cite{jrz2}. However, the
state considered in ref.~\cite{jrz2} has $J^P=1^-$ with no definite
charge conjugation. A molecular state with a vector and a scalar $D_s$ mesons
with negative charge conjugation was studied by the first time in 
ref.~\cite{rapha}, and the obtained mass was $(4.42\pm0.10)~\GeV$, also 
consistent with the $X(4350)$ mass, but with not consistent quantum numbers.
A molecular state with a vector and a scalar $D_s$ mesons
with positive charge conjugation can be constructed using the combination
$D_s^{*+}{D}_{s0}^{*-}-D_s^{*-}D_{s0}^{*+}$.

There is already some interpretations for this state. In ref.~\cite{xliu}
it was interpreted as a excited $P$-wave charmonium state 
$\Xi_{c2}^{\prime\prime}$ and in ref.~\cite{wang} it was interpreted as a
mixed charmonium-$D_s^*D_s^*$ state.
In this work, we use the QCD sum rules (QCDSR) \cite{svz,rry,SNB}, to
study the two-point function based on a  $D_s^*{D}_{s0}^*$ current with 
$J^{PC}=1^{-+}$, to test if the new observed resonance structure, $X(4350)$, 
can be interpreted as such molecular state, as suggested by Belle Coll.
\cite{bellex}.


The  QCD sum rule approach is based on the two-point correlation function
\beq
\Pi_{\mu\nu}(q)=i\int d^4x ~e^{iq.x}\lag 0
|T[j_\mu(x)j_\nu^\dagger(0)]|0\rag,
\lb{2po}
\enq
where  a current that couples with a $J^{PC}=1^{-+}$ $D_s^*{D}_{s0}^*$ 
state is given by:
\beq
j_\mu={1\over\sqrt{2}}\left[(\bar{s}_a\gamma_\mu c_a)(\bar{c}_bs_b)-
(\bar{c}_a\gamma_\mu s_a)(\bar{s}_bc_b)\right]
\;,
\label{field}
\enq
where $a$ and $b$ are color indices. 

Since the current in Eq.~(\ref{field}) is not conserved, we can write
the correlation function in Eq.~(\ref{2po}) in terms of two independent
Lorentz structures:
\beq
\Pi_{\mu\nu}(q)=-\Pi_1(q^2)(g_{\mu\nu}-{q_\mu q_\nu
\over q^2})+\Pi_0(q^2){q_\mu q_\nu\over q^2}.
\lb{lorentz}
\enq
The two invariant functions, $\Pi_1$ and $\Pi_0$, appearing in
Eq.~(\ref{lorentz}), have respectively the quantum numbers of the spin 1
and 0 mesons. Therefore,
we choose to work with the Lorentz structure $g_{\mu\nu}$, since it
gets contributions only from the $1^{-+}$ state.

The QCD sum rule is obtained by evaluating the correlation function in 
Eq.~(\ref{2po}) in two ways: in the OPE side, we
calculate the correlation function at the quark level in terms of
quark and gluon fields.  We work at leading order 
in $\alpha_s$ in the operators, we consider the contributions from 
condensates up to dimension eight and  we keep terms which are linear in
the strange quark mass $m_s$. In the  phenomenological side,
the correlation function is calculated by inserting intermediate states 
for the $D_s^{*}\bar{D}_s^{*}$ molecular scalar state.
Parametrizing the coupling of the exotic state,
$X=D_s^{*}{D}_{s0}^{*}$, to the current, $j_\mu$, in Eq.~(\ref{field}) in 
terms of the parameter $\lambda$:
\beq\label{eq: decay}
\lag 0 |
j_\mu|X\rag =\lambda\varepsilon_\mu.
\label{lam}
\enq
the phenomenological side of Eq.~(\ref{2po}),
in the $g_{\mu\nu}$ structure, can be written as 
\beq
\Pi_1^{phen}(q^2)={\lambda^2\over
M_X^2-q^2}+\int_{0}^\infty ds\, {\rho^{cont}(s)\over s-q^2}, \lb{phe} 
\enq
where the second term in the RHS of Eq.(\ref{phe}) denotes higher
resonance contributions.

The correlation function in the OPE side can be written as a
dispersion relation:
\beq
\Pi_1^{OPE}(q^2)=\int_{4m_c^2}^\infty ds {\rho^{OPE}(s)\over s-q^2}\;,
\lb{ope}
\enq
where $\rho^{OPE}(s)$ is given by the imaginary part of the
correlation function: $\pi \rho^{OPE}(s)=\mbox{Im}[-\Pi_1^{OPE}(s)]$.

As usual in the QCD sum rules method, it is
assumed that the continuum contribution to the spectral density,
$\rho^{cont}(s)$ in Eq.~(\ref{phe}), vanishes bellow a certain continuum
threshold $s_0$. Above this threshold, it is given by
the result obtained with the OPE. Therefore, one uses the ansatz \cite{io1}
\beq
\rho^{cont}(s)=\rho^{OPE}(s)\Theta(s-s_0)\;,
\enq

To improve the matching between the two sides of the sum rule, we 
perfom a Borel transform. After transferring the continuum contribution to 
the OPE side, the sum rules for the exotic meson, described by a $1^{-+}$ 
$D_s^*{D}_{s0}^*$  molecular current,
up to dimension-eight condensates, using factorization hypothesis, can
be written as:
\beq \lambda^2e^{-\mme^2/M^2}=\int_{4m_c^2}^{s_0}ds~
e^{-s/M^2}~\rho^{OPE}(s)\;, \lb{sr}
\label{sr1}
\enq
where
\beq
\rho^{OPE}(s)=\rho^{pert}(s)+\rh^{\qqs}(s)+\rh^{\lag G^2\rag}(s)
+\rh^{mix}(s)+\rh^{\qqs^2}(s)+\rh^{\lag 8 \rag}(s),
\lb{rhoeq}
\enq
with
\begin{eqnarray*}
\label{rhoope}
&&\rho^{pert}(s)={1\over 2^{12} \pi^6}\int\limits_{\almi}^{\alma}
{d\al\over\alpha^3}
\int\limits_{\bemi}^{1-\al}{d\be\over\be^3}(1-\al-\be)
F^3(\al,\be) \left[ 3(1+\al+\be)F(\al,\be) + 2m_c^2(1-\al-\be)^2 \right], \\
&& \rho^{m_{s}}(s)=-{3m_{s}m_{c}\over 2^{9} \pi^6}\int\limits_{\almi}^{\alma}
{d\al\over\alpha^3} \int\limits_{\bemi}^{1-\al}{d\be\over\be^2}(1-\al-\be)^{2}
F^3(\al,\be) , \\
&&\rho^{\qqs}(s)=-{3m_{c}\qqs\over2^{6} \pi^4} 
\int\limits_{\almi}^{\alma} {d\al\over\alpha^2}
\int\limits_{\bemi}^{1-\al}{d\be\over\be} (1-\al-\be)F^{2}(\al,\be), \\
&&\rho^{m_{s}\cdot \qqs}(s)=-{3m_{s}\qqs\over2^{7} \pi^4} \left[
\int\limits_{\almi}^{\alma} {d\al\over\alpha}
\int\limits_{\bemi}^{1-\al}{d\be\over\be}
m_{c}^{2} (3+\al+\be) F(\al,\be) -
 \int\limits_{\almi}^{\alma} {d\al\over\alpha(1-\al)} H^{2}(\al) \right], \\
&&\rho^{\lag G^2\rag}(s)=-{\Gd\over3\:2^{12}\pi^6}
\int\limits_{\almi}^{\alma} {d\al\over\al^3}
\int\limits_{\bemi}^{1-\al}{d\be\over\be}\left[
6\al(1-2\al-2\be) F^{2}(\al,\be) -  \right. \\&&\\
&& \hspace{3cm} \left.
-3m^{2}_{c}(1-\al-\be) \left\{ 1+\al(1-2\al) +\be(\al+3\be)\right\} F(\al,\be) -
m_{c}^{4} \be(1-\al-\be)^{3} \right], \\
&&\rho^{mix}(s)={3m_c\mixs\over2^{8} \pi^4}
\int\limits_{\almi}^{\alma} {d\al\over\al^2}
\int\limits_{\bemi}^{1-\al}{d\be\over\be} \left[ 
\al(1-\al) -\be(5\al+2\be) \right] F(\al,\be), \\
&&\rho^{ms\cdot mix}(s)={m_s\mixs\over2^{8} \pi^4} \left[
\int\limits_{\almi}^{\alma} {d\al\over\al}
\int\limits_{\bemi}^{1-\al}{d\be} \left\{ 
m_{c}^{2}(3+5\al+4\be) - \al\be s \right\} - \right. \\
&& \hspace{7cm} \left.
-\int\limits_{\almi}^{\alma} {d\al\over\al} \left\{
m_{c}^{2}(2+\al) - \al(1-\al)s (2-7\al) \right\} \right], \\
&&\rho^{\qqs^2}(s)=-{\qqs^2\over 2^{6}3\pi^2}(8m_c^2+s)\sqrt{1-4m_c^2/s} 
,\\
&&\rho^{m_{s}\cdot \qqs^2}(s)=-{m_{c} m_{s} \qqs^2\over 2^{5}\pi^2}
\sqrt{1-4m_c^2/s},\\ 
&&\rho^{\lag 8 \rag}(s)={m_c^2\qqs \mixs \over2^{6}\pi^2}{\sqrt{1-4m_c^2/s}
\over s},\\
&&\Pi^{\lag 8 \rag}(M^{2})={m_{c}^2 \qqs \mixs \over 2^6 \pi^2}
\int\limits_{0}^{1} {d\al\over(1-\al)} ~e^{- {m_{c}^{2} \over \al(1-\al)
M^{2}} } \left[
{1 - 3\al} + {2 m_c^2\over \al M^2} \right], \\
&&\Pi^{m_{s}\cdot \lag 8 \rag}(M^{2})={m_{s} m_{c} \qqs \mixs \over 3\: 2^7 
\pi^2}
\int\limits_{0}^{1} {d\al\over\al} ~e^{- {m_{c}^{2} \over \al(1-\al)M^{2}} } 
\left[{m_{c}^{2}\over M^{2}}{(6-4\al-10\al^{2})\over \al(1 - \al)} - 
(6-13\al+20\al^{2}) \right], 
\end{eqnarray*}
where we use the following definitions:
\beq
F(\al,\be) = m_{c}^{2}(\al+\be) - \al\be s,
\enq
\beq
H(\al) = m_{c}^{2}-\al(1-\al)s.
\enq
The integration limits are given by $\almi=({1-\sqrt{1-
4m_c^2/s})/2}$, $\alma=({1+\sqrt{1-4m_c^2/s})/2}$, $\bemi={\al
m_c^2/( s\al-m_c^2)}$.
We have neglected the contribution of the dimension-six
condensate $\langle g^3 G^3\rangle$, since it is assumed to be suppressed
by the loop factor $1/16\pi^2$.  

To extract the mass $\mme$ we take the derivative of Eq.~(\ref{sr})
with respect to $1/M^2$, and divide the result by Eq.~(\ref{sr}).

For a consistent comparison with the results obtained for the other molecular
states using the QCDSR approach, we  have considered here the same values 
used for the quark masses and condensates  as in 
refs.~\cite{x3872,molecule,lee,bracco,rapha,z12,zwid,narpdg}:
$m_c(m_c)=(1.23\pm 0.05)\,\GeV $, $m_s=(0.13\pm 0.03)\,\GeV $,
$\lag\bar{q}q\rag=\,-(0.23\pm0.03)^3\,\GeV^3$, $\qqs=0.8\qq$,
$\lag\bar{s}g\si.Gs\rag=m_0^2\lag\bar{s}s\rag$ with $m_0^2=0.8\,\GeV^2$,
$\lag g^2G^2\rag=0.88~\GeV^4$.

\begin{figure}[h]
\centerline{\epsfig{figure=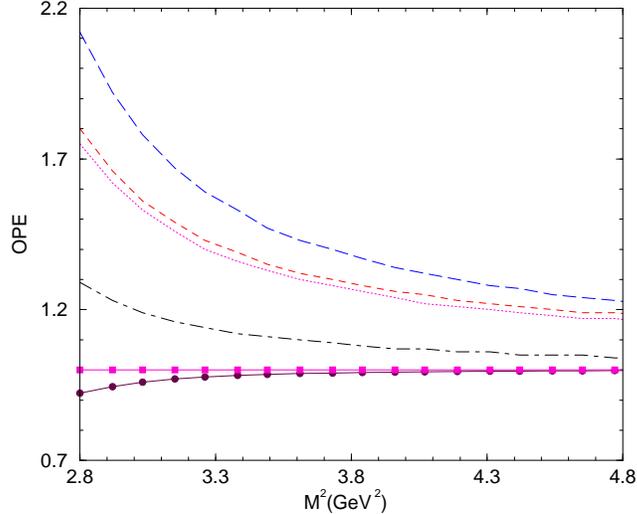,height=70mm}}
\caption{The OPE convergence for the $J^{PC}=1^{-+},~D_s^*D_{s0}^*$ 
molecule in the region
$2.8 \leq M^2 \leq4.8~\GeV^2$ for $\sqrt{s_0} = 5.3$ GeV.  We plot the 
relative contributions starting with the perturbative contribution plus de 
$m_s$ correction (long-dashed line), and each other line represents the 
relative contribution after adding of one extra condensate in the expansion: 
+ $\qqs+m_s\qqs$  (dashed line), 
+ $\langle g^2G^2\rangle$ (dotted line), + $\mixs+m_s\mixs$ 
(dot-dashed line), + $\qqs^2+m_s\qqs^2$ (line with circles), + $<8>+m_s<8>$ 
(line with squares).}
\label{figconv}
\end{figure}

The Borel window is determined by analysing the OPE convergence, the Borel
stability  and the pole contribution. To determine the minimum value of the 
Borel mass we impose that
the contribution of the dimension-8 condensate should be smaller than 20\% of 
the total contribution. 

\begin{figure}[h]
\centerline{\epsfig{figure=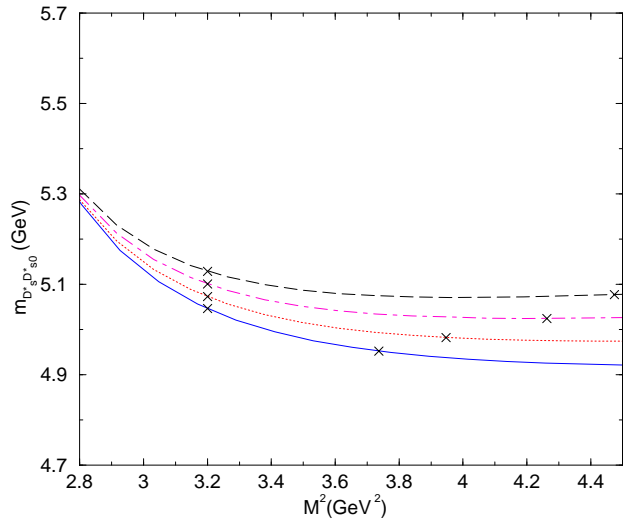,height=70mm}}
\caption{The exotic meson mass, described with a $D_s^*D_{s0}^*$ molecular 
current, as a function of the sum rule parameter
($M^2$) for $\sqrt{s_0} =5.3$ GeV (solid line), $\sqrt{s_0} =5.4$ GeV (dotted 
line), $\sqrt{s_0} =5.5$ GeV (dot-dashed line),
and $\sqrt{s_0} =5.6$ GeV (dashed line). The crosses
indicate the upper and lower limits in the Borel region.}
\label{figmx}
\end{figure}

In Fig.~\ref{figconv} we show the contribution of all the terms in the
OPE side of the sum rule. From this figure we see that for $M^2\geq 2.8$ 
GeV$^2$ the contribution of the dimension-8 condensate is less than 10\% of the
total contribution, which indicates a good Borel convergence. However, from
 Fig.~\ref{figmx} we see that the Borel stability is good only for $M^2\geq 
3.2$ GeV$^2$. Therefore, we  fix the lower
value of $M^2$ in the sum rule window as $M^2_{min}= 3.2$ GeV$^2$.

The maximum value of the Borel mass is 
determined by imposing that the pole contribution must be bigger than the 
continuum contribution. 

\begin{figure}[h]
\centerline{\epsfig{figure=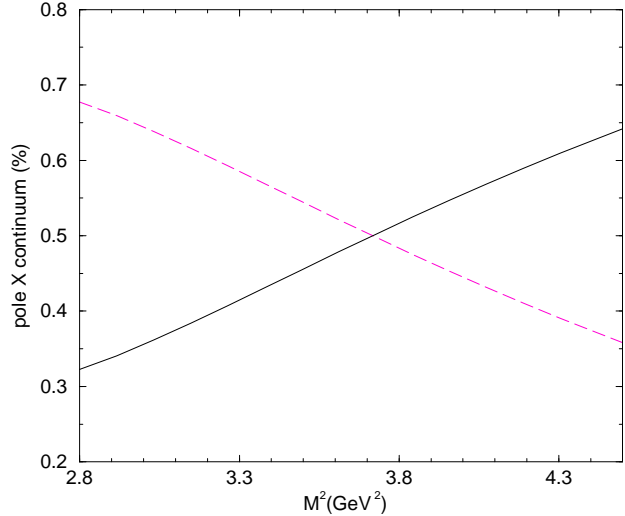,height=70mm}}
\caption{The dashed line shows the relative pole contribution (the
pole contribution divided by the total, pole plus continuum,
contribution) and the solid line shows the relative continuum
contribution for $\sqrt{s_0}=5.3~\GeV$.}
\label{figpvc}
\end{figure}

From Fig.~\ref{figpvc} we see that for $\sqrt{s_0}=5.3~\GeV$, the pole 
contribution is bigger than the continuum contribution for $M^2\leq3.74\GeV^2$.
We show in Table I the values of $M_{max}$ for other values of $\sqrt{s_0}$.
For $\sqrt{s_0}\leq5.1~\GeV$ there is no allowed Borel window.

\begin{center}
\small{{\bf Table I:} Upper limits in the Borel window for the $!^{-+},~
D_s^*D_{s0}^*$ current 
obtained from the sum rule for different values of $\sqrt{s_0}$.}
\\
\begin{tabular}{|c|c|}  \hline
$\sqrt{s_0}~(\GeV)$ & $M^2_{max}(\GeV^2)$  \\
\hline
 5.2 & 3.42 \\
\hline
 5.3 & 3.74 \\
\hline
5.4 & 3.95 \\
\hline
5.5 & 4.26 \\
\hline
5.6 & 4.47 \\
\hline
\end{tabular}\end{center}

Using the Borel window, for each value of $s_0$, to evaluate the mass of the
exotic meson and then varying the value of the continuum threshold in
 the range $5.3\leq \sqrt{s_0} \leq5.6$
GeV, we get $\mme = (5.04\pm0.09)~\GeV$.

Up to now we have kept the values of the quark masses and condensates fixed.
To check the dependence of our results with these values we fix 
$\sqrt{s_0}=5.45~\GeV$ and vary the other parameters
in the ranges: $m_c=(1.23\pm0.05)~\GeV$, $m_s=(0.13\pm 0.03)\,\GeV $,
$\lag\bar{q}q\rag=\,-(0.23\pm0.03)^3\,\GeV^3$, $m_0^2=(0.8\pm0.1)\GeV^2$. 
In our calculation we have assumed the factorization hypothesis. However, it 
is important to check how a violation of the factorization hypothesis would
modify our results. For this reason we multiply $\qqs^2$  and $\qqs\mixs$
in Eq.~(\ref{rhoope}) by a factor $K$ and we vary $K$ in the range 
$0.5\leq K\leq2$. We notice that the results are more sensitive to the
variations on the values of $\qq$ and $K$.

Taking into account the uncertainties given above we get
\beq
\mme = (5.05\pm0.19)~\GeV,
\label{ymass}
\enq
which is not compatible with the mass of the narrow structure $X(4350)$
observed by Belle. It is, however, very interesting to notice that the mass
obtained for a state described with a $1^{--},~ D_s^*D_{s0}^*$ molecular
current is $m_{1^{--}}=(4.42\pm0.10)~\GeV$, much smaller than what we have
obtained with the $1^{-+},~ D_s^*D_{s0}^*$ molecular current. This may be 
interpreted as an indication that it is easier to form molecular states with
not exotic quantum numbers.

\begin{figure}[h]
\centerline{\epsfig{figure=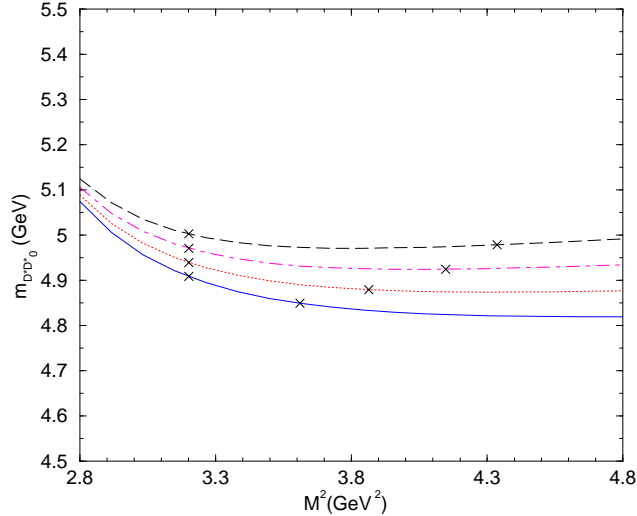,height=70mm}}
\caption{The $1^{-+}$ meson mass, described with a $D^*D_{0}^*$ molecular 
current, as a function of the sum rule parameter
for $\sqrt{s_0} =5.2$ GeV (solid line), $\sqrt{s_0} =5.3$ GeV (dotted 
line), $\sqrt{s_0} =5.4$ GeV (dot-dashed line),
and $\sqrt{s_0} =5.5$ GeV (dashed line). The crosses
indicate the upper and lower limits in the Borel region.}
\label{masx0}
\end{figure}

From the above study it is very easy to get results for the $D^*{D}_0^*$ 
molecular type current with $J^{PC}=1^{-+}$. For this we only have to take
$m_s=0$ and $\sss=\qq$ in Eq.~(\ref{rhoope}). The OPE convergence in this case
is very similar to the preliminar case, and we also get a good Borel
stability only for $M^2\geq3.2~\GeV^2$. Fixing $M^2_{min}=3.2~\GeV^2$, the 
minimum allowed value for the continuum, thresold is $\sqrt{s_0}=5.2~\GeV$.
We show, in Fig.~\ref{masx0}, the result for the mass of such state using 
different values of the continuum threshold, with the upper and lower limits
in the Borel region indicated.

Using the  values of the
continuum threshold in the range $5.2\leq \sqrt{s_0}\leq5.5~\GeV$
we get for the state described with a  $1^{-+},~D^*D_{0}^*$ molecular 
current: $m_{D^*D_0^*}=(4.92\pm0.08)~\GeV$. Approximately one hundred MeV 
bellow the value obtained for the similar strange state. In the case of the
$D^*D_{0}^*$ molecular  current with $J^{PC}=1^{--}$, the mass obtained was
\cite{rapha}: $m_{1^{--}}=(4.27\pm0.10)~\GeV$, again much smaller than for 
the exotic case.

In conclusion, we have presented a QCDSR analysis of the two-point
function based on  $D_s^*{D}_{s0}^*$ and $D^*{D}_0^*$ molecular 
type currents with $J^{PC}=1^{-+}$. Our findings indicate that the $X(4350)$ 
narrow structure observed by  the Belle Collaboration
in the process $\gamma\gamma\to X(4350)\to J/\psi\phi$, cannot be described
by using a exotic $1^{-+},~D_s^*{D}_{s0}^*$ current.

\section*{Acknowledgements}

 This work has been partly supported by FAPESP and CNPq-Brazil.


\end{document}